\documentclass[conference,compsoc]{IEEEtran}

\ifCLASSOPTIONcompsoc
  \usepackage[nocompress]{cite}
\else
  \usepackage{cite}  % normal IEEE
\fi

\usepackage{multirow}
\usepackage{balance}
\usepackage[pdftex]{graphicx}
\graphicspath{{./figures/}{./jpeg/}} % declare path(s) where your graphic files are
\DeclareGraphicsExtensions{.pdf,.jpeg,.png}

\usepackage{url} % Basically, \url{my_url_here}.

\begin{document} % Titles are generally capitalized 
%\title{Detection of extremist users and prediction of susceptible targets and adopters of radical content on social media}
%\title{Forecasting the dynamics of extremist propaganda campaigns on social media}
%\title{Forecasting extremist propaganda campaigns on social media}
\title{Predicting online extremism, content adopters, and interaction reciprocity}
% information diffusion of social media campaigns for extremist propaganda 

\author{\IEEEauthorblockN{Emilio Ferrara\IEEEauthorrefmark{1},
Wen-Qiang Wang\IEEEauthorrefmark{1},
Onur Varol\IEEEauthorrefmark{2}, 
Alessandro Flammini\IEEEauthorrefmark{2} and
Aram Galstyan\IEEEauthorrefmark{1}}
\IEEEauthorblockA{\IEEEauthorrefmark{1}Information Sciences Institute, University of Southern California, Marina del Rey, CA 90292, USA}
\IEEEauthorblockA{\IEEEauthorrefmark{2}School of Informatics and Computing, Indiana University, Bloomington, IN 47401, USA}
emiliofe@usc.edu, wenqianw@usc.edu, ovarol@indiana.edu, aflammin@indiana.edu, galstyan@isi.edu}

% author names and affiliations % use a multiple column layout for up to three 
%\author{
%\IEEEauthorblockN{Emilio Ferrara}
%\IEEEauthorblockA{Information Sciences Institute\\
%University of Southern California\\
%Marina del Rey, CA 90292\\
%emiliofe@usc.edu}
%\and
%\IEEEauthorblockN{Wenqiang Wang}
%\IEEEauthorblockA{Information Sciences Institute\\
%University of Southern California\\
%Marina del Rey, CA 90292\\
%wenqianw@usc.edu}
%\and
%\IEEEauthorblockN{Onur Varol}
%\IEEEauthorblockA{School of Informatics and Computing\\
%Indiana University\\
%Bloomington, IN 47401\\
%ovarol@indiana.edu}
%\and
%\IEEEauthorblockN{Alessandro Flammini}
%\IEEEauthorblockA{School of Informatics and Computing\\
%Indiana University\\
%Bloomington, IN 47401\\
%aflammin@indiana.edu}
%\and
%\IEEEauthorblockN{Aram Galstyan}
%\IEEEauthorblockA{Information Sciences Institute\\
%University of Southern California\\
%Marina del Rey, CA 90292\\
%galstyan@isi.edu}
%}

\maketitle

% As a general rule, do not put math, special symbols or citations
% in the abstract
\begin{abstract}
We present a machine learning framework that leverages a mixture of metadata, network, and temporal features to detect extremist users, and predict content adopters and interaction reciprocity in social media.
We exploit a unique dataset containing millions of tweets generated by more than 25 thousand users who have been manually identified, reported, and suspended by Twitter due to their involvement with extremist campaigns. 
We also leverage millions of tweets generated by a random sample of 25 thousand regular users who were exposed to, or consumed, extremist content.
We carry out three forecasting tasks, (i) to detect extremist users, (ii) to estimate whether regular users will adopt extremist content, and finally (iii) to predict whether users will reciprocate contacts initiated by extremists. 
All forecasting tasks are set up in two scenarios: a \textit{post hoc} (time independent) prediction task on aggregated data, and a simulated real-time prediction task. The performance of our framework is extremely promising, yielding in the different forecasting scenarios up to 93\% AUC for extremist user detection, up to 80\% AUC for content adoption prediction, and finally up to 72\% AUC for interaction reciprocity forecasting.
We conclude by providing a thorough feature analysis that helps determine which are the emerging signals that provide predictive power in  different scenarios.
\end{abstract}

% no keywords

% For peerreview papers, this IEEEtran command inserts a page break and
% creates the second title. It will be ignored for other modes.
\IEEEpeerreviewmaketitle

\section{Introduction}

Researchers are devoting increasing attention to the issues related to online extremism, terrorist propaganda and radicalization campaigns \cite{schiermeier2015terrorism, reardon2015terrorism}.
Social media play a central role in these endeavors, as increasing evidence from social science research suggests \cite{berger2013who, fisher2015jihadist}. 
For example, a widespread consensus on the relationship between social media usage and the rise of extremist groups like the Islamic State of Iraq and al-Sham (viz. ISIS) has emerged among policymakers and security experts \cite{stern2015isis, cockburn2015rise, weiss2015isis}. ISIS' success in increasing its roster to thousands of members has been related in part to a savvy use of social media for propaganda and recruitment purposes. 
One reason is that, until recently, social media platform like Twitter provided a public venue where single individuals, interest groups, or organizations, were given the ability to carry out extremist discussions and terrorist recruitment, without any form of restrictions, and with the possibility of gathering audiences of potentially millions. 
Only recently, some mechanisms have been put into place, based on manual reporting, to limit these abusive forms of communications.
Based on this evidence, we argue in favor of developing computational tools capable of effectively analyzing massive social data streams, to detect extremist users, to predict who will become involved in interactions with radicalized users, and finally to determine who is likely to consume extremist content.
The goal of this article is to address the three questions above by proposing a computational framework for detection and prediction of extremism in social media. 
We tapped into Twitter to obtain a relevant dataset, leveraged expert crowd-sourcing for annotation purposes, and then designed, trained and tested the performance of our prediction system in static and simulated real-time forecasts, as detailed below.

\subsection*{Contributions of this work}
\noindent The main contributions of our work can be summarized as: 

\begin{itemize}
\item We formalize three different forecasting tasks related to online extremism, namely the detection of extremist users, the prediction of adoption of extremist content, and the forecasting of interaction reciprocity between regular users and extremists.
\item We propose a machine prediction framework that analyzes social media data and generates features across multiple dimensions, including user metadata, network statistics, and temporal patterns of activity, to perform the three forecasting tasks above.
\item We leverage an unprecedented dataset that contains over 3 millions tweets generated by over 25 thousand extremist accounts, who have been manually identified, reported, and suspended by Twitter. We also use around 30 million tweets generated by a random sample of 25 thousand regular users who were exposed to, or consumed, extremist content. 
\item For each forecasting task, we design two variants: a post-hoc (time independent) prediction task performed on aggregated data, and a simulated real-time forecast where the learning models are trained as if data were available up to a certain point in time, and the system must generate predictions on the future.
\item We conclude our analysis by studying the predictive power of the different features employed for prediction, to determine their role in the three forecasts.
\end{itemize}

\section{Data and Preliminary Analysis}
In this section we describe our dataset, the curation strategy yielding the annotations, and some preliminary analysis.

\subsection{Sample selection and curation}
In this work we rely on data and labels constructed by using a procedure of manual curation and expert verification. 
We retrieved on a public Website a list of over 25 thousands Twitter accounts whose activity was labeled as supportive of the Islamic State by the crowd-sourcing initiative called \emph{Lucky Troll Club}. 
The goal of this project was to leverage annotators with expertise in Arabic languages to identify ISIS accounts and report them to Twitter. Twitter's anti-abuse team manually verifies all suspension requests, and grants some based on the active violation of Twitter's Terms of Service policy against terrorist- or extremist-related activity.
Here we focus on the 25,538 accounts that have been all suspended between January and June 2015 by Twitter as a consequence of evidence of activity supporting the Islamic State group. For each account, we also have at our disposal information about the suspension date, and the number of followers of that user as of the suspension date.

\subsection{Twitter data collection}
The next step of our study consisted in collecting data related to the activity of the 25,538 ISIS supporters on Twitter. To this purpose, we leveraged the Twitter \textit{gardenhose} data source (roughly 10\% of the Twitter stream) collected by Indiana University~\cite{davis2016botornot}. We decided to collect not only the tweets generated by these accounts prior to their suspension, but also to build a dataset of their targets. In particular, we are concerned with accounts unrelated to ISIS with whom the ISIS supporters tried to establish some forms of interaction. We therefore constructed the following two datasets:

\smallskip

\noindent{\textbf{ISIS accounts}}: this dataset contains 3,395,901 tweets generated in the time interval January-June 2015 by the 25,538 accounts identified by Twitter as supporters of ISIS. This is a significant portion of all the accounts suspended by Twitter in relation to ISIS.\footnote{The Guardian recently reported that between April 2015 and February 2016, Twitter's anti-terror task force suspended about 125,000 accounts linked to ISIS extremists: \url{http://www.theguardian.com/technology/2016/feb/05/twitter-deletes-isis-accounts-terrorism-online}} 

\smallskip

\noindent{\textbf{Users exposed to ISIS}}: this dataset contains 29,193,267 tweets generated during January-June 2015 by a set of 25 thousand users randomly sampled among the larger set of users that constitute ISIS accounts' followers.
This set is by choice of equal size to the former one, to avoid introducing class imbalance issues. 

\smallskip

For prediction purposes, we will use as positive and negative labels the \textit{ISIS accounts} group and the accounts in the \textit{users exposed to ISIS}, respectively.

\section{Methodology}
In this section we discuss the learning models and the features adopted by our framework. The complete prediction pipeline (learning models, cross validation, feature selection, and performance evaluation) is developed using Python and the scikit-learn library~\cite{pedregosa2011scikit}.

\subsection{Learning models}
We adopt two off-the-shelf learning models as a proof of concept for the three classification tasks that we will discuss later (see~\S\ref{sec:experiments}): \textit{Logistic Regression} and \textit{Random Forests}.

\smallskip

\noindent{\textbf{Logistic Regression}}: The first implemented algorithm is a simple Logistic Regression (LR) with LASSO regularization. The advantage of this approach is its scalability, which makes it very effective to (possibly real-time) classification tasks on large datasets. The only parameter to tune is the loss function \textit{C}. We expect that LR will provide the baseline classification and prediction performance.

\smallskip

\noindent{\textbf{Random Forests}}: We also use a state-of-the-art implementation of \emph{Random Forests} (RF)~\cite{Breiman2001RF}.
The vectors fed into the learning models represent each user's features. 
\textit{Random Forests} are trained using 100 estimators and adopting the Gini coefficient to measure the quality of splits. Optimal parameters setting is obtained via cross validation (see~\ref{sub:cv}).

\smallskip

Note that the goal of this work is not to provide new  machine learning techniques, but to illustrate that existing methods can provide promising  results. We also explored  additional learning models (e.g., SVM, Stochastic Gradient Descent, etc.), which provide comparable prediction performance but are less computationally efficient and scalable.

\subsubsection{Cross validation} \label{sub:cv}
The results of our performance evaluation (see~\S\ref{sec:experiments}) are all obtained via $k$-fold cross validation. We adopt $k=5$ folds, and therefore use 80\% of data for training, and the remainder 20\% for testing purpose, averaging performance scores across the 5 folds.
We also use 5-fold cross validation to optimize the parameters of the two learning algorithms (LR and RF), by means of an exhaustive cross-validated grid search on the hyperparameter space.

\subsubsection{Evaluation scores} \label{sub:eval}
We benchmark the performance of our system by using four standard prediction quality measures, namely Precision, Recall, F1 (harmonic mean of Precision and Recall), and AUC---short for Area Under the Receiver Operating Characteristic (ROC) curve~\cite{hastie2005elements}.

\begin{table}[!t]
\centering 
\caption{List of 52 features extracted by our framework} 
%\begin{tabular}{@{}l@{}l@{}}
\begin{tabular}{@{} ll @{}}
\hline\multirow{13}{*}{\rotatebox{90}{\textbf{User metadata \& activity}}}
& Number of followers\\
& Number of friends (i.e., followees)\\
& Number of posted tweets\\
& Number of favorite tweets\\
& Ratio of retweets / tweets\\
& Ratio of mentions / tweets\\
& Avg number of hashtags\\
& (avg, var) number of retweets\\
& Avg. number of mentions\\
& Avg. number of mentions (excluding retweets)\\
& Number of URLs in profile description\\
& (avg, std, min, max, proportion) URLs in tweets\\
& Length of username\\

\hline\multirow{4}{*}{\rotatebox{90}{\textbf{Timing}}} 
%\hline\multirow{4}{*}{\textbf{Timing}}
& (avg, var) number of tweets per day\\
& (avg, std, min, max) interval between two consecutive tweets\\
& (avg, std, min, max) interval between two consecutive retweets\\
& (avg, std, min, max) interval between two consecutive mentions\\

\hline\multirow{5}{*}{\rotatebox{90}{\textbf{Netw. stats}}} 
& (avg, std, min, max) distribution of retweeters' number of followers\\
& (avg, std, min, max) distribution of retweeters' number of friends\\
& (avg, std, min, max) distribution of mentioners' number of followers \\
& (avg, std, min, max) distribution of mentioners' number of friends \\
& (avg, std, min, max) number of retweets of the tweets  by others\\
\hline
\end{tabular} 
\label{tab:features}
\end{table}

\subsection{Feature engineering and feature selection}
We manually crafted a set of 52 features belonging to three different classes: user metadata, timing features, and network statistics, as detailed below.

\smallskip

\noindent{\textbf{User metadata and activity features:}}
User metadata have been proved pivotal to model classes of users in social media~\cite{mislove2011understanding, ferrara2014rise}.
We build user-based features leveraging the metadata provided by the Twitter API related to the author of each tweet, as well as the source of each retweet. User features include the number of tweets, followers and friends associated to each users, the frequency of adoption of hashtags, mentions, and URLs, and finally some profile descriptors. In total, 18 user metadata and activity features are computed (cf. Table~\ref{tab:features}).

\smallskip

\noindent{\textbf{Timing features:}}
Important insights may be concealed by the temporal dimension of content production and consumption, as illustrated by recent work~\cite{Ghosh11snakdd, subrahmanian2016darpa}. A basic timing feature is the average number of tweets posted per day. Other timing features include statistics (average, standard deviation, minimum, maximum) of the intervals between two consecutive events, e.g., two tweets, retweets, or mentions. Our framework generates  14 timing features (cf. Table~\ref{tab:features}). 

\smallskip

\noindent{\textbf{Network statistics:}}
Twitter content spreads from person to person via retweets and mentions. We expect that the emerging network structure carries important information to characterize different types of communication. Prior work shows that using network features significantly helps prediction tasks like social bot detection~\cite{ferrara2014rise, davis2016botornot, subrahmanian2016darpa}, and campaign detection~\cite{ratkiewicz2011detecting, ferrara2016detection}. Our framework focuses on two types of networks: (i) retweet, and (ii) mention networks. Users are nodes of such networks, and retweets or mentions are directed links between pairs of users. For each user, our framework computes the distribution of followers and friends of all users who retweet and mention that user, and extracts some descriptive statistics (average, standard deviation, minimum, maximum) of these distributions. Our system builds  20 network statistics features (cf. Table~\ref{tab:features}).

\subsubsection{Greedy  feature selection}
Our framework generates a set $F$ of $|F|=52$ features.
In our type of prediction tasks, some features  exhibit more predictive power than others: temporal dependencies introduce strong correlations among some features, thus some possible redundancy. Among the different existing ways to select the most relevant features for the prediction task at hand~\cite{guyon2003introduction}, in the interest of computational efficiency, we adopted a simple  greedy forward feature selection method,  as follows: (i)~initialize the set of selected features $S = \emptyset$; (ii)~for each feature $f \in F-S$, consider the union set $U = S \cup f$; (iii)~train the classifier using the features in $U$; (iv)~test the average performance of the classifier trained on this set; (v)~add to $S$ the feature that provides the best performance; (vi)~repeat (ii)-(v) as long as a significant performance increase  is yield.

\section{Experiments}\label{sec:experiments}
In the following, we formalize three different prediction problems related to online extremism: 

\smallskip

\noindent{\textbf{Task I (T1): Detection of extremist supporters.}} The first task that our system will face is a binary classification aimed to detect ISIS accounts and separate them from those of regular users. The problem is to test whether any predictive signal is present in the set of features we designed to characterize social media activity related to extremism, and serves as a yardstick for the next two more complex problems.

\smallskip

\noindent{\textbf{Task II (T2): Predicting extremist content adoption.}} The set of 25 thousand users we randomly sampled among followers of ISIS accounts can be leveraged to perform the prediction of extremist content adoption. We define as a positive instance of adoption in this context when a regular user retweets some content s/he is exposed to that is generated by an ISIS account. 

\smallskip

\noindent{\textbf{Task III (T3): Predicting interactions with extremists.}} The third task presents likely the most difficult challenge: predicting whether a regular user will engage into interactions with extremists. A positive instance of interaction is represented by a regular user replying to a contact initiated by an ISIS account.

\smallskip

\noindent{\textbf{Static versus real-time predictions.}} For each of the three prediction tasks described above, we identified two modalities, namely a static (time independent) and a simulated real-time prediction. In the former scenario, a static prediction ignores temporal dependencies in that the system aggregates all data available across the entire time range (January-June 2015), and then performs training and testing using the 5-fold cross validation strategy by randomly splitting datapoints into the 5 folds and averaging the prediction performance across folds. In the latter scenario, a real-time prediction is simulated in which data are processed for training and testing purposes by respecting the timeline of content availability: for example, the system can exploit the first month of available data (January 2015) for training, and then producing predictions for the remainder 5 months (Feb-Jun 2015), for which performance is tested.

The performance of our framework in the three tasks, each with the two prediction modalities,  is discussed in the following. The section concludes with the analysis of feature predictive power (see~\S\ref{sub:feature_analysis}).

\begin{figure*}[!t]
\includegraphics[width=.87\columnwidth]{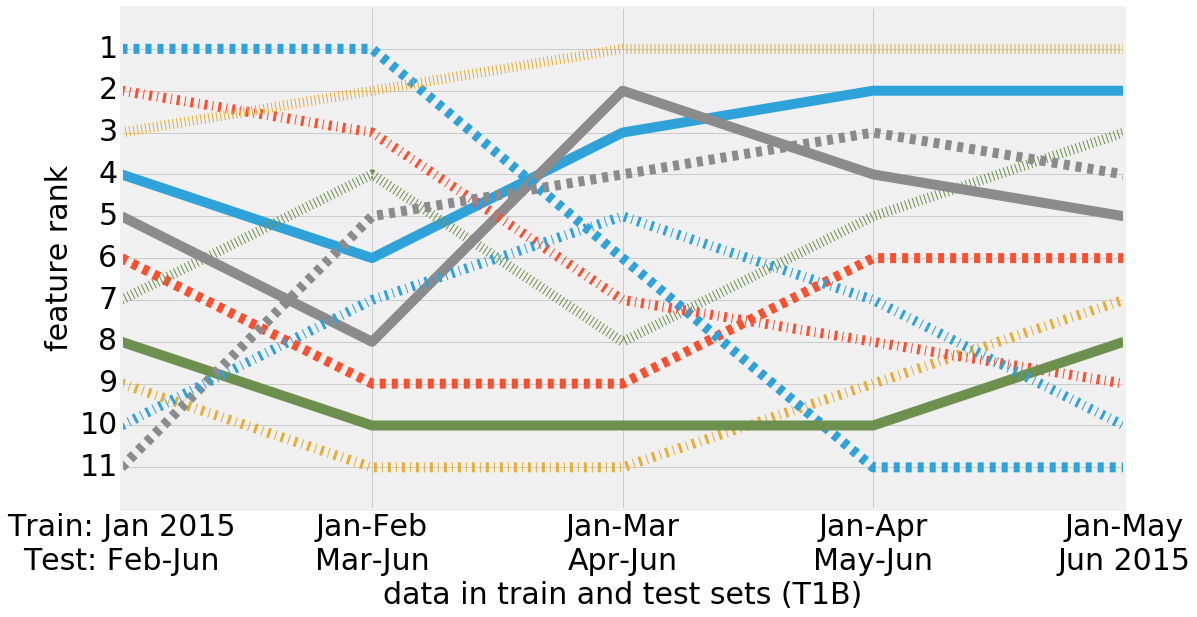}
\includegraphics[width=1.2\columnwidth]{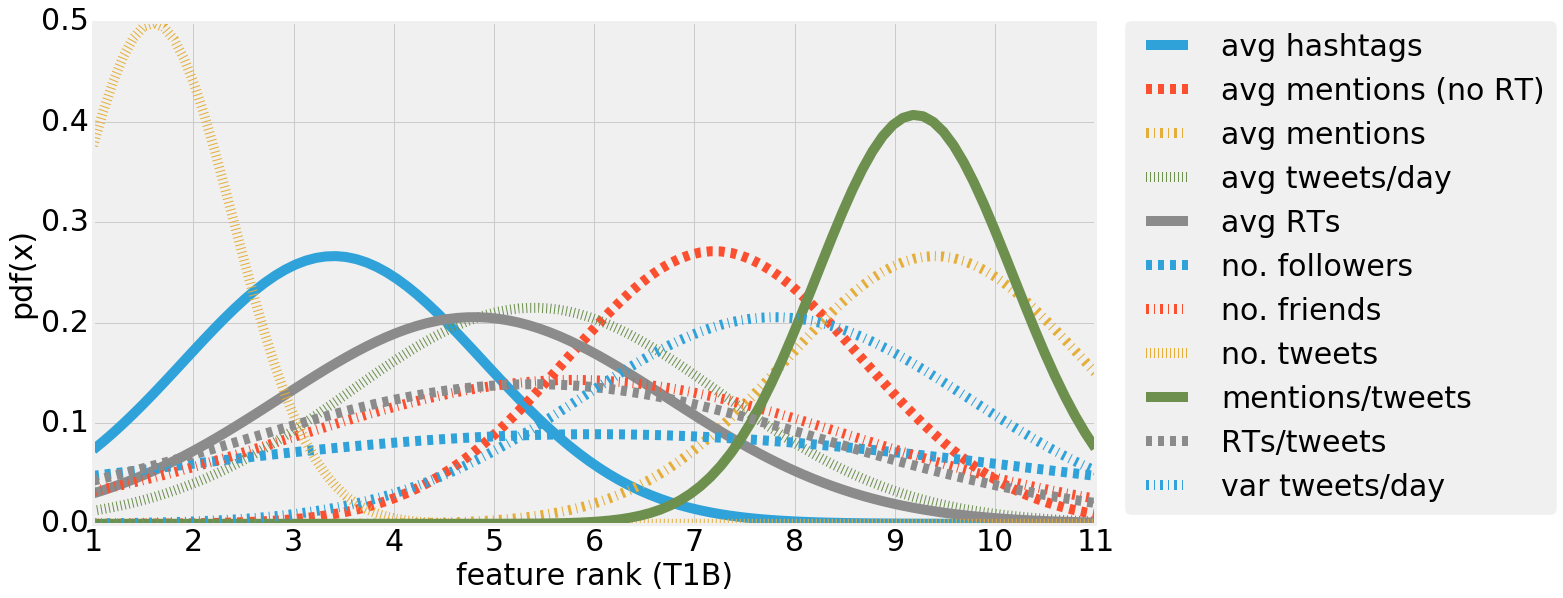}
\caption{T1B: Feature selection analysis and feature rank distribution (top 11 features)}
\label{fig:t1b}
\end{figure*}

\subsection{T1: Detection of extremist supporters}
In the following we discuss the static (T1A) and real-time (T1B) scenarios for the first prediction task, namely detecting extremist accounts on Twitter.

\subsubsection{T1A: Time-independent detection}
The detection of extremist user accounts is the most natural task to start the performance evaluation of our framework. 
Our analysis aims at verifying that the 52 features we carefully hand-crafted indeed carry some predictive signal useful to separate extremist users from regular ones. The dataset at hand contains two roughly equal-sized classes (about 25 thousand instances each), where ISIS accounts are labeled as positive instances, and regular users as negative ones. Each instance is a characterized by a 52-dimensional vector, and positive and negative examples are fed to the two learning models (LR and RF). The first task, in short T1A, is agnostic of time dependencies: data are aggregated throughout the entire 6 months period (January-June 2015) and training/testing is performed in a traditional 5-fold cross-validated fashion (cf.~\S\ref{sub:cv}). 
Table~\ref{tab:t1a} summarizes the performance of LR and RF according to the four quality measures  described above (cf.~\ref{sub:eval}): Both models perform  well, with \textit{Random Forests} achieving an accuracy above 87\% as measured by AUC. These results are encouraging and  demonstrate that simple off-the-shelf models can yield good performance in T1A. 

%\begin{table}[!ht] \centering
%\caption{T1A Time-aggregated data (11 features)}
%\begin{tabular}{@{}l l l l l @{}}
%& Precision & Recall & F1 & AUC \\
%\hline\hline
%Logistic Regression & 0.817 & 0.289 & 0.427 & 0.735 \\
%Random Forests & 0.780 & 0.862 & 0.819 & 0.812 \\
%\hline\hline
%\end{tabular}
%\label{tab:t1A12}
%\end{table}

\begin{table}[!h] \centering\vspace*{-2.5mm}
\caption{Extremists  detection (T1A)}
\begin{tabular}{@{}l l l l l @{}}
& Precision & Recall & F1 & AUC \\
\hline
Logistic Regression & 0.778 & 0.506 & 0.599 & 0.756 \\
Random Forests & 0.855 & 0.893 & 0.874 & \textbf{0.871} \\
\hline
\end{tabular}
\label{tab:t1a}\vspace*{-2.5mm}
\end{table}

\subsubsection{T1B: Simulated real-time detection} A more complex variant of this prediction task is by taking into account the temporal dimension. Our prior work has demonstrated that accounting for temporal dependencies is very valuable in social media prediction tasks and significantly improves prediction performance~\cite{ferrara2016detection}: therefore we expect that the performance of our framework in a simulated real-time prediction task will exceed that of the static scenario.

In this simulated real-time prediction task, T1B, we divide the available data into temporal slices used separately for training and prediction purposes. Table~\ref{tab:t1b} reports five columns, each of which defines a scenario where one or more months of data are aggregated for training, and the rest is used for prediction and performance evaluation. For example, in the first column, the learning models are trained on data from January 2015, and the prediction are performed and evaluated on future data in the interval February-June 2015.\\

\textit{Random Forests} greatly benefits from accounting for temporal dependencies in the data, and the prediction performance as measured by AUC ranges between 83.8\% (with just one month of training data) to an excellent 93.2\% (with five months of training data).
Fig.~\ref{fig:t1b}(left) illustrates the ranking of the top 11 features identified by feature selection, as a function of the number of months of data in the training set. Fig.~\ref{fig:t1b}(right) displays the distributions of the rankings of each feature across the 5 different temporal slices. For the extremist users detection task, the most predictive features are (1) number of tweets, (2) average number of hashtags, and (3) average number of retweets. One hypothesis is that extremist users are more active than average users, and therefore exhibit distinctive patterns related to volume and frequency of activity.
% Feature analysis will be  discussed further in \S~\ref{sub:feature_analysis}.

\begin{table}[!h] \centering\vspace*{-2.5mm}
\caption{Real-time extremists  detection (T1B)}
\begin{tabular}{@{}l l l l l l @{}}
Training: & Jan & Jan-Feb & Jan-Mar & Jan-Apr & Jan-May \\
Testing: & Feb-Jun & Mar-Jun & Apr-Jun & May-Jun & Jun \\
\hline 
AUC (LR) &  0.743 & 0.753 & 0.655 & 0.612 & 0.602 \\
Precision (LR) & 0.476 & 0.532 & 0.792 & 0.816 & 0.796 \\
Recall (LR) & 0.629 & 0.675 & 0.377 & 0.289 & 0.275 \\
F1 (LR) & 0.542 & 0.595 & 0.511 & 0.427 & 0.409 \\
\hline 
AUC (RF) & \textbf{0.838} & \textbf{0.858} & \textbf{0.791} & \textbf{0.942} & \textbf{0.932} \\
Precision (RF) & 0.984 & 0.922 & 0.868 & 0.931 & 0.910 \\
Recall (RF) & 0.679 & 0.733 & 0.649 & 0.957 & 0.959 \\
F1 (RF) & 0.804 & 0.817 & 0.743 & 0.944 & 0.934 \\
\hline 
\end{tabular}
\label{tab:t1b}\vspace*{-2.5mm}
\end{table}

\begin{figure*}[!t]
\includegraphics[width=.87\columnwidth]{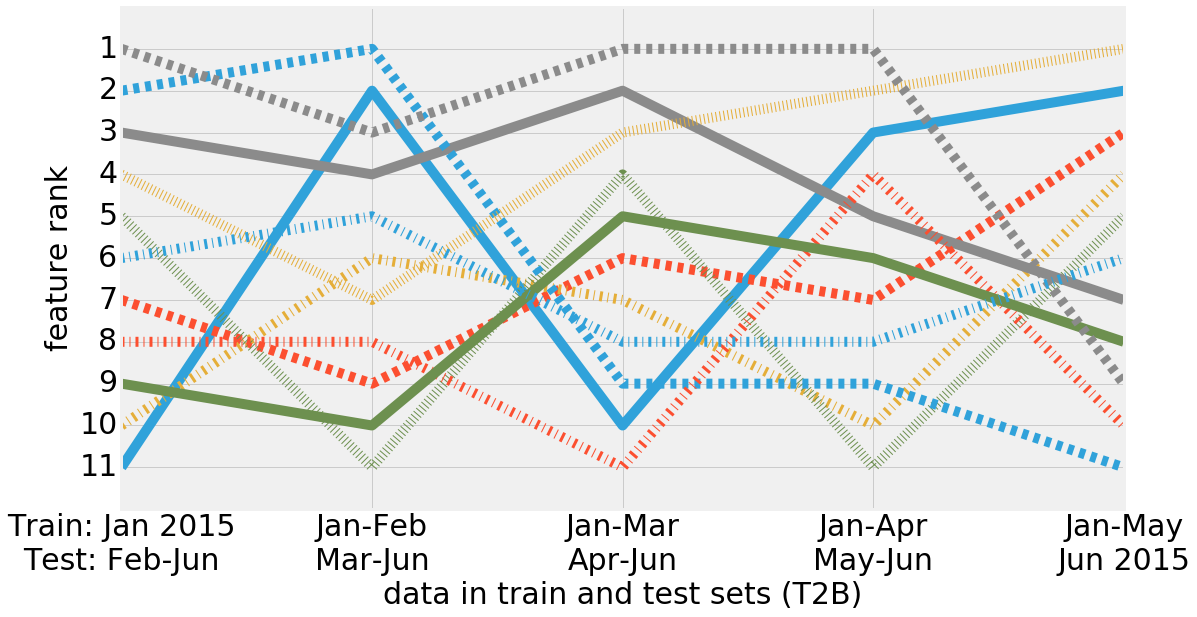}
\includegraphics[width=1.2\columnwidth]{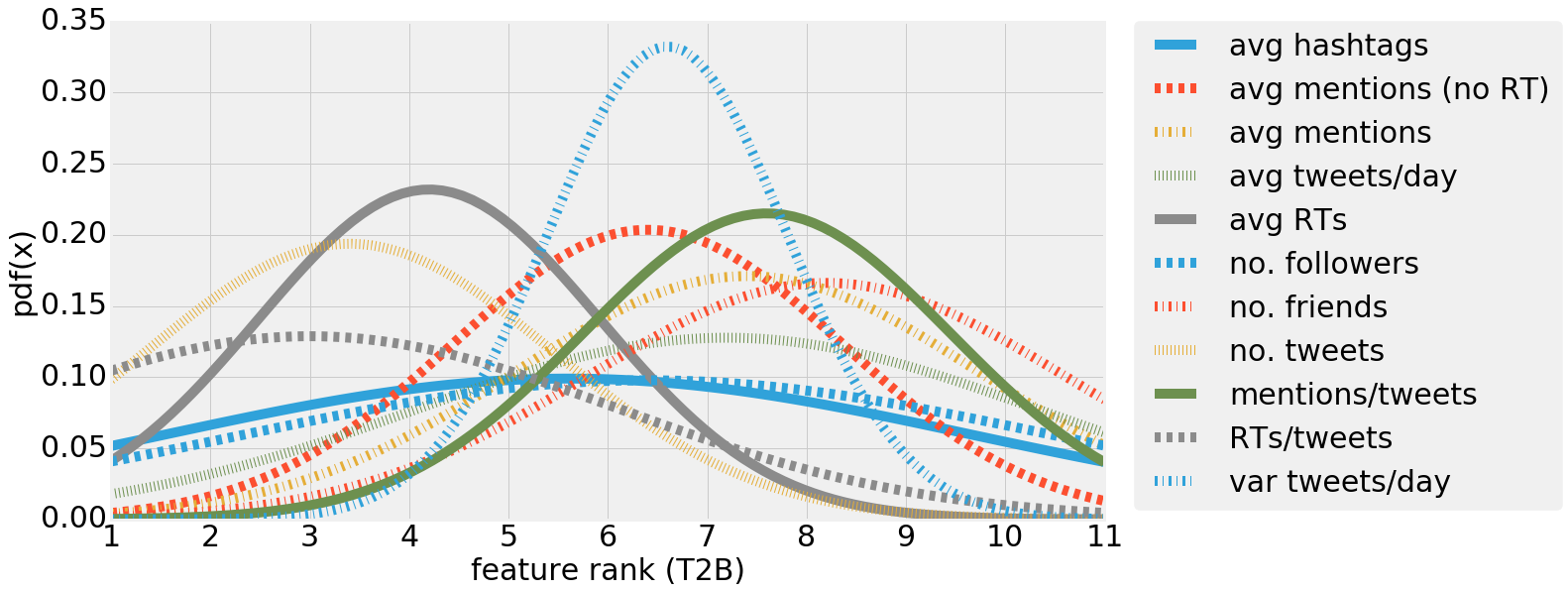}
\caption{T2B:  Feature selection analysis and feature rank distribution (top 11 features)}
\label{fig:t2b}
\end{figure*}

\subsection{T2: Predicting extremist content adoption}
The second task, namely predicting the adoption of extremist content by regular users, is discussed in the static (T2A) and real-time (T2B) scenarios in the following.

\subsubsection{T2A: Time-independent prediction}
The first instance of T2 is again on the time-aggregated datasets spanning January-June 2015. 
Predicting content adoption is a known challenging task, and a wealth of literature has explored the factors behind online information contagion~\cite{lerman2010information}.
In this scenario, we aim to predict whether a regular user will retweet a content produced by an ISIS account. Positive instances are represented by users who retweeted at least one ISIS tweet in the aggregated time period, while negative ones are all users exposed to such tweets who did not retweet any of them.
Table~\ref{tab:t2a} summarizes the performance of our models: \textit{Random Forests} emerges again as the best performer, although the gap with \textit{Logistic Regression} is narrow, and the latter provides significantly better Recall. Overall, T2A appears clearly more challenging than T1A, as the top performance yields a 77.1\% AUC score. The results on the static prediction are promising and set a good baseline for the real-time prediction scenario, discussed next.

\begin{table}[!ht] \centering\vspace*{-2.5mm}
\caption{Content adoption prediction (T2A)}
\begin{tabular}{@{}l l l l l @{}}
& Precision & Recall & F1 & AUC \\
\hline 
Logistic Regression & 0.433  & 0.813 & 0.565 & 0.755 \\
Random Forests & 0.745 & 0.615 & 0.674 & \textbf{0.771} \\
\hline 
\end{tabular}
\label{tab:t2a}\vspace*{-2.5mm}
\end{table}

\subsubsection{T2B: Simulated real-time prediction}
We again consider temporal data dependencies to simulate a real-time prediction for T2. 
Similarly to T1B, in T2B we preserve the temporal ordering of data, and divide the dataset in training and testing according to month-long temporal slices, as summarized by Table~\ref{tab:t2b}.
\textit{Random Forests} again seems to benefit from the temporal correlations in the data, and the prediction performance at peak improves up to 80.2\% AUC. \textit{Logistic Regression} fails again at exploiting temporal information, showing some performance deterioration if compared to T2A.
Fig.~\ref{fig:t2b} shows that, for the content adoption prediction, the ranking of the top 11 features in T2B is less stable than that of T1B. The top three most predictive features for this task are (1) ratio of retweets over tweets, (2) number of tweets, and (3) average number of retweets. Note that the latter two top features also appear in the top 3 of the previous task, suggesting an emerging pattern of feature predictive dynamics.

\begin{table}[!ht] \centering%\vspace*{-2.5mm}
\caption{Real-time content adoption prediction (T2B) }
\begin{tabular}{@{}l l l l l l @{}}
Training: & Jan & Jan-Feb & Jan-Mar & Jan-Apr & Jan-May \\
Testing: & Feb-Jun & Mar-Jun & Apr-Jun & May-Jun & Jun \\
\hline 
AUC (LR) & 0.682  & 0.674  & 0.673  & 0.703  & 0.718 \\
Precision (LR) & 0.188 & 0.240 & 0.148 & 0.116 & 0.043 \\
Recall (LR) & 0.814 & 0.367 & 0.345 & 0.725 & 0.362 \\
F1 (LR) & 0.305 & 0.290 & 0.207 & 0.199 & 0.077 \\
\hline 
AUC (RF)  & \textbf{0.565} & \textbf{0.598} & \textbf{0.676} & \textbf{0.779} & \textbf{0.802} \\
Precision (RF) & 0.433 & 0.384 & 0.266 & 0.205 & 0.070 \\
Recall (RF) & 0.087 & 0.070 &  0.336 & 0.648  & 0.813 \\
F1 (RF) & 0.145 & 0.119 & 0.297 & 0.311 & 0.130 \\
\hline 
\end{tabular}
\label{tab:t2b}\vspace*{-2.5mm}
\end{table}

\begin{figure*}[!t]
\includegraphics[width=.87\columnwidth]{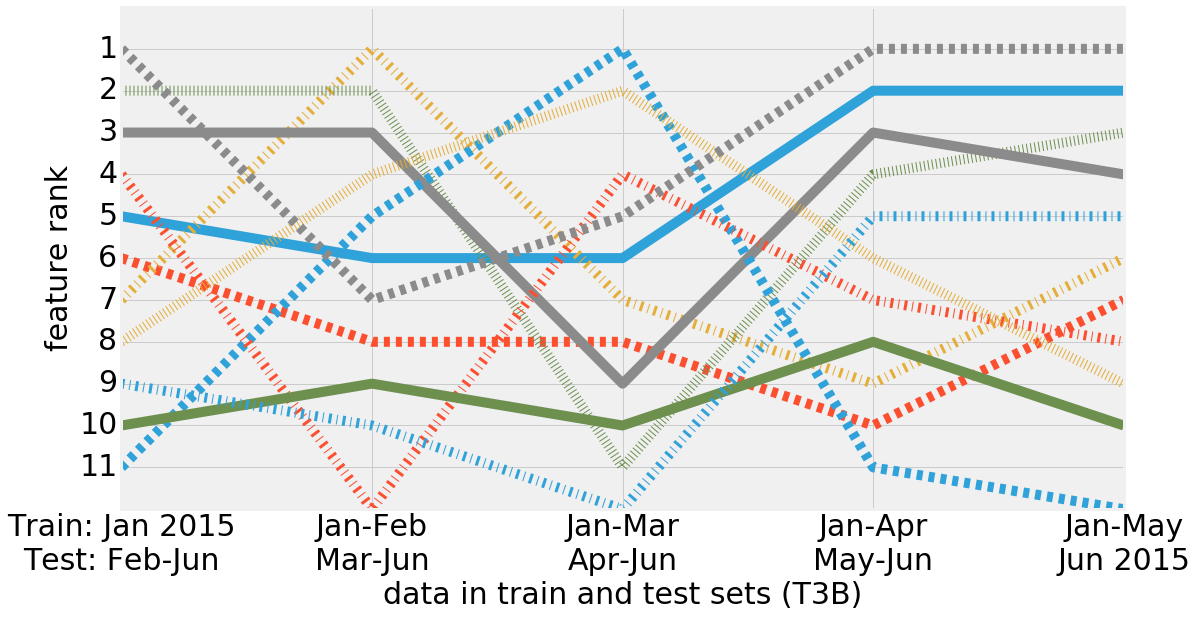}
\includegraphics[width=1.2\columnwidth]{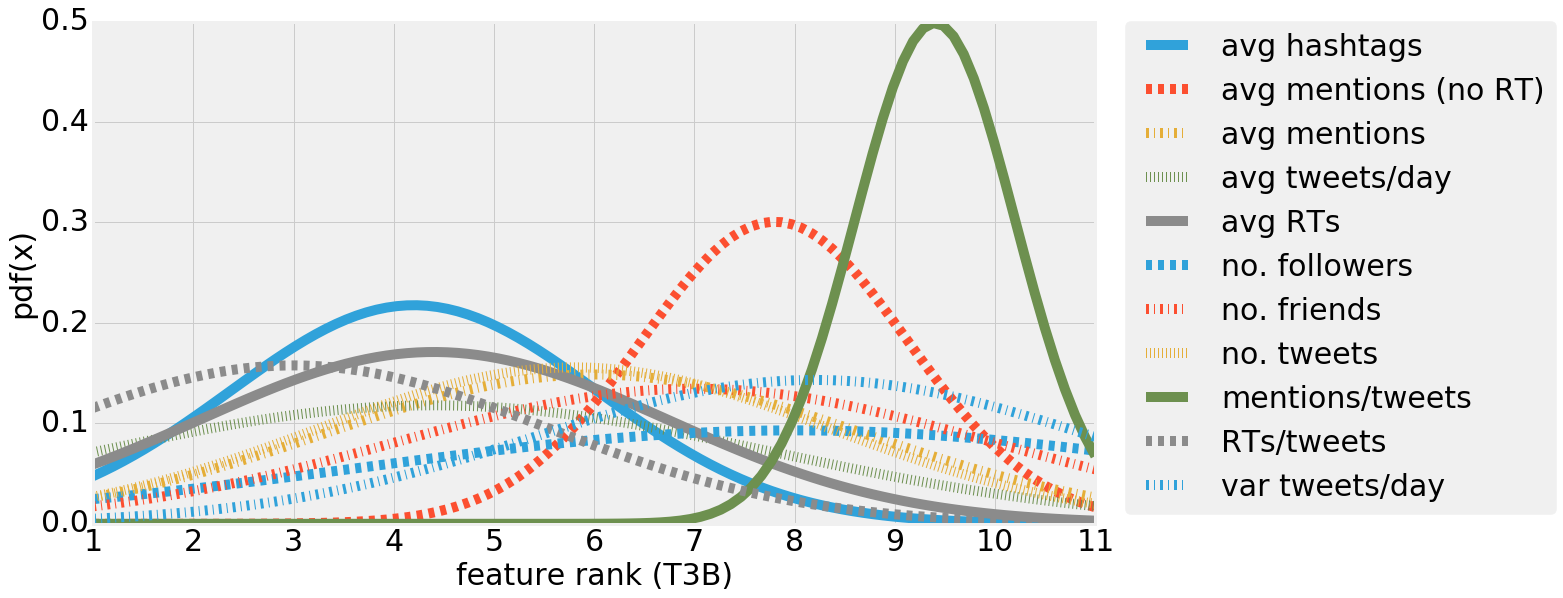}
\caption{T3B: Feature selection analysis and feature rank distribution (top 11 features)}
\label{fig:t3b}
\end{figure*}

\subsection{T3: Predicting interactions  with extremists}
Our third and last task, namely the prediction of interactions between regular users and extremists, is discussed in the following, again separately for the static (T3A) and real-time (T3B) scenarios.

\subsubsection{T3A: Time-independent prediction}
We expect the interaction prediction task to be the most challenging among the three tasks we proposed. Similarly to content adoption prediction, recent literature has explored the daunting challenge of predicting interaction reciprocity and intensity in social media \cite{gilbert2009predicting}.
Consistently with the prior two tasks, our first approach to interaction prediction is time agnostic: we plan to test whether our system is capable to predict whether a regular user who is mentioned by an extremist account will reply back or not. In this case, positive instances are represented by users who reply to at least one contact initiated by ISIS in the aggregated time period (January-June 2015), whereas negative instances are those regular users who did not reply to any ISIS contact.
Table~\ref{tab:t3a} reports the prediction performance of our two models: overall, the task proves challenging as expected, being \textit{Random Forests} the best performer, yielding excellent Recall and 69.2\% AUC. \textit{Logistic Regression} provides fair performance with a 65.8\% AUC score, and both Precision and Recall around 69\%.

\begin{table}[!ht] \centering\vspace*{-2.5mm}
\caption{Interaction reciprocity prediction (T3A)}
\begin{tabular}{@{}l l l l l @{}}
& Precision & Recall & F1 & AUC \\
\hline 
Logistic Regression & 0.697 & 0.690 & 0.693 & 0.658 \\
Random Forests & 0.686 & 0.830 & 0.751  & \textbf{0.692} \\
\hline
\end{tabular}
\label{tab:t3a}\vspace*{-2.5mm}
\end{table}

\subsubsection{T3B: Simulated real-time prediction} 
The final task discussed in this paper is the interaction prediction with temporal data.
Given the complexity of this problem, as demonstrated by  T3A, we plan to test whether incorporating the temporal dimension will help our models achieve better performance. 
Table~\ref{tab:t3b} shows that this appears to be the case: \textit{Random Forests} exhibits an improved temporal prediction performance, boasting up to 72.6\% AUC, using the first 5 months of data for traning, and the last month for prediction and evaluation. \textit{Logistic Regression} improves as well, jumping to a 68.3\% AUC score. 
Both models provide very good Precision/Recall performance, if one considers the challenging nature of predicting interaction reciprocity within our context.
Fig.~\ref{fig:t3b} summarizes the top 11 features ranking, this time showing a more clear division among top features. The top three features in the interaction reciprocity prediction are (1) ratio of retweets over tweets, (2) average number of hashtags, and (3) average number of retweets. Note that all three features already occurred in the top features of the two previous tasks, reinforcing the notion of a clear pattern of feature predictive power, discussed next.

\begin{table}[!ht] \centering\vspace*{-2.5mm}
\caption{Real-time interaction reciprocity prediction (T3B) }
\begin{tabular}{@{}l l l l l l @{}}
Training: & Jan & Jan-Feb & Jan-Mar & Jan-Apr & Jan-May \\
Testing: & Feb-Jun & Mar-Jun & Apr-Jun & May-Jun & Jun \\
\hline 
AUC (LR) & 0.610  & 0.589  & 0.618  & 0.638  & 0.683 \\
Precision (LR) & 0.562 & 0.560 & 0.574 & 0.553 & 0.367 \\
Recall (LR) & 0.720 & 0.775 & 0.813 & 0.783 & 0.647 \\
F1 (LR) & 0.631 & 0.650 & 0.672 & 0.649 & 0.468 \\
\hline 
AUC (RF) & \textbf{0.628} & \textbf{0.633} & \textbf{0.649} & \textbf{0.671} & \textbf{0.726} \\
Precision (RF) & 0.614 & 0.627 & 0.603 & 0.637 & 0.542 \\ 
Recall (RF) & 0.779 & 0.676 & 0.641 & 0.717 & 0.765 \\
F1 (RF) & 0.687 & 0.650 & 0.621 & 0.675 & 0.634 \\
\hline 
\end{tabular}
\label{tab:t3b}\vspace*{-2.5mm}
\end{table}

\subsection{Feature predictive-power analysis} \label{sub:feature_analysis}
We conclude our analysis by discussing the predictive power of the features adopted by our framework.
First, the choice to focus on the top 11 features, rather than the more traditional top 10, is justified by the occurrence of two \emph{ex aequo} in the final ranking of top features, displayed in Table~\ref{tab:feature_ranking}.
Here, we report the ranking of the top 11 features in the three tasks  above. Feature selection is performed on the real-time prediction tasks (not on the time-aggregated ones). This analysis captures the essence of the predictive value of our hand-crafted features in the context of real-time predictions. A clear pattern emerges: (1) ratio of retweets over tweets, (2) average number of hashtags, (2 \textit{ex-aequo}) number of tweets, and (4) average number of retweets, consistently ranked in the top  features for the three different prediction tasks. This insight is encouraging: all these features can be easily computed from the metadata reported by the Twitter API, and therefore could be potentially implemented in a real-time detection and prediction system operating on the Twitter stream with unparalleled efficiency.

\begin{table}[!h] \centering%\vspace*{-2.5mm}
\caption{Feature ranking across the 3 prediction tasks}
\begin{tabular}{@{} l@{} r r r | r@{}}
Feature & Rank: T1B & T2B & T3B & Final \\
\hline 
Ratio of retweets / tweets & 4 & 1 & 1 & \textbf{1} \\
Avg number of hashtags & 2 & 4 & 2 & \textbf{2} \\
Number of tweets & 1 & 2 & 5 & \textbf{=} \\
Avg number of retweets & 3  & 3 & 3 & \textbf{4} \\
Avg tweets per day & 5 & 8 & 4 & \textbf{5} \\
Avg no. mentions (w/out retweets) & 8 & 5 & 8 & \textbf{6} \\
Number of followers & 7 & 6 & 9 & \textbf{7} \\
Number of friends & 6 & 11 & 7 & \textbf{8} \\
Avg number of mentions & 11 & 9 & 6 & \textbf{9} \\
Var tweets / day & 9 & 7 & 10 & \textbf{=} \\
Ratio of mentions / tweets & 10 & 10 & 11 & \textbf{11} \\
\hline 
\end{tabular}
\label{tab:feature_ranking}\vspace*{-2.5mm}
\end{table}

\section{Related Literature}

Two relevant research trends recently emerged in the \textit{computational social sciences} and in the \textit{computer science} communities, discussed separately in the following.\\
\textbf{Computational social sciences research}. This research line is concerned more with understanding the social phenomena revolving around extremist propaganda using online data as a proxy to study individual and group behaviors.
Various recent studies focus on English- and Arabic-speaking audiences online to study the effect of ISIS' propaganda and radicalization.
One example of the former  is the work by Berger and collaborators that provided quantitative evidence of ISIS' use of social media for propaganda. In a 2015 study \cite{berger2015isis}, the authors characterized the Twitter population of ISIS supporters, quantifying its size, provenance, and organization. They argued that most of ISIS' success on Twitter is due to a restricted number of highly-active accounts (500-1000 users). Our analysis illustrates that indeed a limited number of ISIS accounts achieved a very high visibility and followership.
Berger's subsequent work \cite{berger2016isis} however suggested that ISIS' reach (at least among English speakers) has stalled for months as of the beginning of 2016, due to more aggressive account suspension policies enacted by Twitter. Again, a limited amount of English accounts sympathetic to ISIS was found (less than one thousand), and these users were mostly interacting with each other, while being only marginally successful at acquiring other users' attention. This analysis suggests a mechanism of diminishing returns for extremist social media propaganda. 

Using Twitter data as a historical archive, some researchers \cite{magdy2016failedrevolutions} recently tried to unveil the roots of support for ISIS among the Arabic-speaking population. Their analysis seems to suggest that supporters of the extremist group have been discussing about Arab Spring uprisings in the past significantly more than those who oppose ISIS on Twitter. Although their method to separate ISIS supporters from opposers is simplistic, the findings relating narrative framing and recruitment mechanisms are compatible with the literature on social protest phenomena \cite{gonzalez2011dynamics, conover2013geospatial, conover2013digital}.

A few studies explored alternative data sources: one interesting example is the work by Vergani and Bliuc \cite{vergani2015evolution} that uses sentiment analysis (Linguistic Inquiry and Word Count \cite{tausczik2010psychological}) to investigate how language evolved across the first 11 Issues of Dabiq, the flagship ISIS propaganda magazine. Their analysis offers some insights about ISIS radicalization motives, emotions and concerns. For example, the authors found that ISIS has become increasingly concerned with females, reflecting their need to attract women to create their utopia society, not revolving around warriors but around families. ISIS also seems to have increased the use of internet jargon, possibly to connect with the identities of young individuals online. \\
\textbf{Computer science research}. This research stream concerns more with the machine learning and data aspects, to model, detect, and/or predict social phenomena such as extremism or radicalization often with newly-developed techniques.

One of the first computational frameworks, proposed by Bermingham \emph{et al.}~\cite{bermingham2009combining} in 2009, combined social network analysis with sentiment detection tools to study the agenda of a radical YouTube group: the authors examined the topics discussed within the group and their polarity, to model individuals' behavior and spot signs of extremism and intolerance, seemingly more prominent among female users. The detection of extremist content (on the Web) was also the focus of a 2010 work by Qi \emph{et al.}~\cite{qi2010hierarchical}. The authors applied hierarchical clustering to extremist Web pages to divide them into different categories (religious, politics, etc.).

Scanlon and Gerber proposed the first method to detect cyber-recruitment efforts in 2014~\cite{scanlon2014automatic}. They exploited data retrieved from the Dark Web Portal Project~\cite{chen2008uncovering}, a repository of posts compiled from 28 different online fora on extremist religious discussions (e.g., Jihadist) translated from Arabic to English. After annotating a sample of posts as recruitment efforts or not, the authors use Bayesian criteria and a set of textual features to classify the rest of the corpus, obtaining good accuracy, and highlighted the most predictive terms.

Along the same trend, Agarwal and Sureka proposed different machine learning strategies~\cite{agarwal2014focused, sureka2014learning, agarwal2015using, agarwal2016spider} aimed at detecting radicalization efforts, cyber recruitment, hate promotion, and extremist support in a variety of online platforms, including YouTube, Twitter and Tumblr. Their frameworks leverage features of contents and metadata, and combinations of crawling and unsupervised clustering methods, to study the online activity of Jihadist groups on the platforms mentioned above.

Concluding, two very recent articles \cite{rowe2016mining, johnson2016new} explore the activity of ISIS on social media. The former~\cite{rowe2016mining} focuses on Twitter and aims at detecting users who exhibit signals of behavioral change in line with radicalization: the authors suggest that out of 154K users only about 700 show significant signs of possible radicalization, and that may be due to social homophily rather than the mere exposure to propaganda content. The latter study~\cite{johnson2016new} explores a set of 196 pro-ISIS aggregates operating on VKontakte (the most popular Russian online social network) and involving about 100K users, to study the dynamics of survival of such groups online: the authors suggest that the development of large and potentially influential pro-ISIS groups can be hindered by targeting and shutting down smaller ones.
For additional pointers we refer the interested reader to two recent literature reviews on this topic~\cite{correa2013solutions, agarwal2015applying}.

\section{Conclusions}
In this article we presented the problem of predicting online extremism in social media. 
We defined three machine learning tasks, namely the detection of extremist users, the prediction of extremist content adoption, and the forecasting of interactions between extremist users and regular users.
We tapped into the power of a crowd-sourcing project that aimed at manually identifying and reporting suspicious or abusive activity related to ISIS radicalization and propaganda agenda, and collected annotations to build a ground-truth of over 25 thousand suspended Twitter accounts. We extracted over three million tweets related to the activity of these accounts in the period of time between January and June 2015. We also randomly identified an equal-sized set of regular users exposed to the extremist content generated by the ISIS accounts, and collected almost 30 million tweets generated by the regular users in the same period of time. 

By means of state-of-the-art learning models we managed to accomplish predictions in two types of scenarios, a static one that ignores temporal dependencies, and a simulated real-time case in which data are processed for training and testing by respecting the timeline of content availability. The two learning models, and the set of 52 features that we carefully crafted, proved very effective in all of the six combinations of forecasts (three prediction tasks each with two prediction modalities, static and real-time). 
The best performance in terms of AUC ranges between 72\% and 93\%, depending on the complexity of the considered task  and the amount of training data available to the system. 

We concluded our analysis by investigating the predictive power of different features. We focused on the top 11 most significant features, and we discovered that some of them, such as the ratio of retweets to tweets, the average number of hashtags adopted, the sheer number of tweets, and the average number of retweets generated by each user, systematically rank very high in terms of predictive power. Our insights shed light on the dynamics of extremist content production as well as some of the network and timing patterns that emerge in this type of online conversation.

Our work is far from concluded: for the future, we plan to identify more realistic and complex prediction tasks, to analyze the network and temporal dynamics of extremist discussion, and to deploy a prototype system that allows for real-time detection of signatures of abuse on social media.

% use section* for acknowledgment
\ifCLASSOPTIONcompsoc
  \section*{Acknowledgments}  % The Computer Society usually uses the plural form
\else
   \section*{Acknowledgment} % regular IEEE prefers the singular form
\fi

This work has been supported in part by the Office of Naval Research (grant no. N15A-020-0053). The funders had no role in study design, data collection and analysis, decision to publish, or preparation of the manuscript. 

% trigger a \newpage just before the given reference
% number - used to balance the columns on the last page
% adjust value as needed - may need to be readjusted if
% the document is modified later
%\IEEEtriggeratref{8}
% The "triggered" command can be changed if desired:
%\IEEEtriggercmd{\enlargethispage{-5in}}

% references section
\bibliographystyle{IEEEtran}
\bibliography{isis_biblio}
\balance

\end{document}